# Concurrence of Quantum Anomalous Hall and Topological Hall Effects in Magnetic Topological Insulator Sandwich Heterostructures

Jue Jiang[1,4], Di Xiao[1,4], Fei Wang[1], Jae-Ho Shin[1], Domenico Andreoli[2], Jianxiao Zhang[1], Run Xiao[1], Yi-Fan Zhao[1], Morteza Kayyalha[1], Ling Zhang[1], Ke Wang[3], Jiadong Zang[2], Chaoxing Liu[1], Nitin Samarth[1], Moses H. W. Chan[1], and Cui-Zu Chang[1]

[1]Department of Physics, The Pennsylvania State University, University Park, PA 16802

[2]Department of Physics, University of New Hampshire, Durham, NH 03824

[3]Materials Research Institute, The Pennsylvania State University, University Park, PA 16802

[4]These authors contributed equally to this work.

Corresponding authors: nxs16@psu.edu (N.S.); mhc2@psu.edu (M.H.W.C.); cxc955@psu.edu (C.Z.C.)

The quantum anomalous Hall (QAH) effect is a quintessential consequence of non-zero Berry curvature in momentum-space. The QAH insulator harbors dissipation-free chiral edge states in the absence of an external magnetic field. On the other hand, the topological Hall (TH) effect, a transport hallmark of the chiral spin textures, is a consequence of real-space Berry curvature. While both the QAH and TH effects have been reported separately, their coexistence, a manifestation of entangled chiral edge states and chiral spin textures, has not been reported. Here, by inserting a TI layer between two magnetic TI layers to form a sandwich heterostructure, we realized a concurrence of the TH effect and the QAH effect through electric field gating. The TH effect is probed by bulk carriers, while the QAH effect is characterized by chiral edge states. The appearance of TH effect in the QAH insulating regime is the consequence of chiral magnetic domain walls that result from the gate-induced

**Dzyaloshinskii-Moriya interaction and occur during the magnetization reversal process in the magnetic TI sandwich samples. The coexistence of chiral edge states and chiral spin textures potentially provides a unique platform for proof-of-concept dissipationless spin-textured spintronic applications.**

Electronic band structures of nontrivial topology in momentum-space and magnetic chiral spin textures in real-space have attracted enormous attention in the past decade since they harbor elegant Berry curvature physics [1, 2, 3]. The intrinsic anomalous Hall (AH) effect is such an example: it is induced by the Berry curvature in momentum-space in ferromagnetic (FM) materials [4] and can even be quantized under certain circumstances, leading to the quantum anomalous Hall (QAH) effect. The QAH effect has been theoretically proposed [5, 6, 7, 8] and experimentally realized [9, 10, 11, 12, 13] in magnetically doped topological insulator (TI) films. On the other hand, chiral spin textures (e.g. skyrmions) provide another example of nontrivial topology, but in real-space. It has been shown that chiral spin textures can also induce a Hall current: this is known as the topological Hall (TH) effect and is generally regarded as the transport signature of non-zero spin chirality [3]. The TH effect has been experimentally observed in many *metallic* systems, such as MnSi [14, 15], MnGe [16], FeGe [17], and SrIrO$_3$/SrRuO$_3$ interface [18, 19] as well as magnetically doped TI films and heterostructures [20, 21]. The TH effect in these systems is usually observed accompanied by the AH effect. However, there is no conclusive evidence to date that the AH effect found in these *metallic* systems to be intrinsic, i.e., exclusively induced by the momentum-space Berry curvature [4].

The QAH and TH effects have been separately observed in magnetically doped TI [9, 10, 20, 21], with distinctly different sample geometries. The QAH effect can be realized only in the *insulating* regime of a magnetic TI [9, 10, 11, 12, 13] while the TH effect is usually seen in *metallic* systems [20, 21]. In this *Article*, we realized the TH effect in the *insulating* regime of magnetic TI and demonstrated the concurrence of the QAH and TH effects in a single sample. This concurrence provides a new platform to understand the interplay between chiral edge states of the QAH effect and chiral spin textures associated with the TH effect.

In order to realize the QAH and TH effects in one TI-based device, the sample needs to accommodate the following three conditions: (*i*) The time-reversal symmetry is broken, which is the common prerequisite for both QAH and TH effects; (*ii*) The chemical potentials of the top and bottom surfaces can be simultaneously tuned near the magnetization exchange gaps, which is essential for the QAH effect; (*iii*) A significant Dzyaloshinskii-Moriya (DM) interaction, $H = \boldsymbol{D}_{ij} \cdot (\boldsymbol{S}_i \times \boldsymbol{S}_j)$, can be created, which is required for the TH effect. Recently, two papers reported the observation of the TH effect in TI. Yasuda *et al* [20] fabricated the Cr-$(Bi,Sb)_2Te_3$/$(Bi,Sb)_2Te_3$ bilayer structures, where only one surface is gapped owing to the magnetic exchange interaction. The consequential spatial asymmetry thus favors the formation of the DM interaction and gives rise to the TH effect in this bilayer structure when it is tuned into the *p*-type *metallic* regime. However, the QAH effect cannot be realized in such a bilayer sample geometry since the other surface is non-magnetic and gapless. Liu *et al* [21] grew 4 quintuple layers (QL) Mn-doped $Bi_2Te_3$ films without a Te capping layer on a $SrTiO_3$ substrate, revealing the TH effect in both *p*- and *n*-type *metallic* regimes. Since the Dirac point is buried under the bulk valence bands in $Bi_2Te_3$ films [22, 23], the QAH state is also unlikely in Mn-doped $Bi_2Te_3$ samples. In uniformly doped QAH samples, the inversion symmetry is preserved in the bulk [9, 10], so that the DM interaction is induced only on the top and bottom surfaces [24, 25]. However, the DM interactions from the two opposite surface states (SSs) have opposite signs. When the two surfaces are strongly coupled, spin chirality cannot be achieved. Effectively, the overall DM interaction felt by the magnetization is greatly reduced and it is difficult to realize the TH effect in such uniformly doped QAH samples. Therefore, to realize the concurrence of the QAH and TH effects in a single sample, the two surfaces of a QAH sample should be separated and inversion symmetry must be broken, in order to induce finite DM interaction.

We fabricated a TI-based sandwich structure with an undoped TI layer (5 QL (Bi, Sb)$_2$Te$_3$ layers) inserted between two magnetic TI layers (two 3 QL Cr-doped (Bi,Sb)$_2$Te$_3$ layers) (**Figs. 1a** and **1b**). Such a 3-5-3 sandwich heterostructure has two distinct advantages. First, the nonmagnetic TI layer serves as a spacer that separates the magnetic exchange interaction between the two magnetic TI layers [26, 27, 28]. As a result, the influence of the DM interaction can be maximized since the magnetic moments in each magnetic TI layer interacts only with their own SS. Second, both the top and bottom SSs are separately gapped by the magnetization, thus making the QAH effect possible. The magnetic/nonmagnetic/magnetic TI sandwich heterostructures were grown on 0.5 mm thick SrTiO$_3$(111) substrates in a molecular beam epitaxy (MBE) chamber with a base vacuum ~ $2\times10^{-10}$ mbar. The transport studies were carried out in a dilution refrigerator (Leiden Cryogenics, 10 mK, 9 T) and a Physical Property Measurement System (Quantum Design, 2 K, 9 T) with the magnetic field applied perpendicular to the film plane. Six-terminal Hall bars with bottom-gate electrodes (**Fig. 1a**) were used for transport studies.

We now focus on the transport results of the 3-5-3 heterostructure. When the bottom gate $V_g = 0$ V, the FM order at low temperatures gaps out the top and bottom SSs, and the chemical potential is located inside the magnetic exchange gaps of both surfaces (**Fig. 1b**). This is confirmed by the observation of a perfectly quantized Hall resistance ($\rho_{yx}$) and a vanishing longitudinal resistance ($\rho_{xx}$) at $T = 30$ mK (**Figs. 1c** and **1d**). With increasing temperature, the sample deviates from the QAH state and shows transport properties of a conventional FM material, namely, hysteretic $\rho_{yx}$ loops and butterfly-shaped $\rho_{xx}$. The Curie temperature ($T_C$) of the 3-5-3 sandwich heterostructure sample is determined to be ~19 K by means of Arrott plots (**Fig. S4**).

**Figure 2** shows the magnetic field $\mu_0H$ dependence of the Hall resistance $\rho_{yx}$ and the longitudinal resistance $\rho_{xx}$ of the 3-5-3 heterostructure at different gate voltages ($V_g - V_g^0$). When $V_g = V_g^0 = +20$ V, the sample displays a perfect QAH state: at zero magnetic field, $\rho_{yx}(0) = \pm h/e^2$ and $\rho_{xx}(0) < 1$ Ω. When the magnetic field $\mu_0H > \mu_0H_c$ ($\mu_0H_c$ is the coercive field), the Hall curves completely overlap during upward and downward magnetic field sweeps (**Fig. 2d**). When ($V_g - V_g^0$) = -100V, hole carriers are injected into the sample and dissipative bulk channels are introduced. In this regime, $\rho_{yx}(0)$ deviates from $h/e^2$ but still remains as high as ~0.76 $h/e^2$ and $\rho_{yx}(0)/\rho_{xx}(0)$ ~1.2, indicating the persistence of the chiral edge transport of the QAH state [9]. The existence of chiral edge transport is further supported by the decrease in $\rho_{xx}(0)$ with decreasing temperature (**Fig. 3e**), which we will discuss in detail below. Notably, over a range of a fraction of a Tesla above $\mu_0H_c$, a "hump" feature appears in the $\rho_{yx}$ curves (green shadow area); specifically, the Hall curve under downward $\mu_0H$ sweep does not overlap with that under upward $\mu_0H$ sweep when $\mu_0H > \mu_0H_c$ (**Fig. 2c**). The "hump" feature observed here has been interpreted as a signature of the TH effect and is considered as strong evidence for the existence of chiral spin textures in real-space [3, 14, 15, 16, 17, 18, 19, 20, 21]. Therefore, by adjusting the chemical potential using the bottom gate, we have realized the concurrence of QAH and TH effects in the *insulating* regime of the magnetic TI sandwich sample. This "hump" feature of TH effect becomes more pronounced when the chemical potential is further tuned towards the bulk valence bands (**Figs. 2a and 2b**).

At ($V_g - V_g^0$) = +70 V, electron carriers are introduced, a trace of the "hump" feature of TH effect also appears (**Fig. 2e**). The observation of $\rho_{yx}(0)$ ~0.85 $h/e^2$ and $\rho_{yx}(0)/\rho_{xx}(0)$ ~ 4.6 demonstrates the existence of the QAH state. Therefore, the QAH effect is also concurrent with the TH effect for $(V_g - V_g^0) = +70V$. When more electrons are introduced, the "hump" feature of TH effect fades away (**Fig. 2f**). The asymmetric behavior of the TH effect in *n*- and *p*-type regions is possibly a result of the non-symmetric electronic band structure of TI [23, 29], which we will discuss below. We note that the slope of the Hall traces at high magnetic fields (0.5 T < $\mu_0 H$ < 1.5 T) is always negative in both $V_g < V_g^0$ and $V_g > V_g^0$ regions, which suggests that the standard Hall coefficient $R_N$ cannot be used to estimate carrier density near the QAH regime (see **Fig. S9** and relevant discussion).

In a FM material, the total $\rho_{yx}$ is a result of three contributions: the normal Hall (NH) resistance $\rho_{yx}^{NH}$, the AH resistance $\rho_{yx}^{AH}$, and the TH resistance $\rho_{yx}^{TH}$,

$$\rho_{yx} = \rho_{yx}^{NH} + \rho_{yx}^{AH} + \rho_{yx}^{TH} \tag{1}$$

In order to single out the TH component $\rho_{yx}^{TH}$, we need to subtract $\rho_{yx}^{NH}$ and $\rho_{yx}^{AH}$ from $\rho_{yx}$. Here we interpret the offset resistance under upward and downward $\mu_0 H$ sweeps (green shadow area in **Fig. 2**) as $\rho_{yx}^{TH}$ for the following reasons. Let us consider the positive $\mu_0 H$ regime: During the downward $\mu_0 H$ sweep (red curves in **Fig. 2**), the system should be in a FM state without any chiral spin textures and thus $\rho_{yx}$ should include $\rho_{yx}^{NH}$ and $\rho_{yx}^{AH}$. For the upward $\mu_0 H$ sweep (blue curves in **Fig. 2**), the system undergoes a magnetic transition around $\mu_0 H_c$ and chiral spin textures, particularly chiral magnetic domain walls (see relevant discussion on **Figs. 4a** and **4b** below), can be formed. In this situation, all three Hall contributions are present and $\rho_{yx}^{NH} + \rho_{yx}^{AH}$ keeps the same value during the downward $\mu_0 H$ sweep. Thus, $\rho_{yx}^{TH}$ can be extracted by the difference between the red and blue curves [21]. We emphasize that the standard expressions for $\rho_{yx}^{NH}$ and $\rho_{yx}^{AH}$ ($\rho_{yx}^{NH} = \mu_0 R_N H$ and $\rho_{yx}^{AH} = R_A M$ with the AH coefficient $R_A$ and the sample magnetization $M$ [9, 30, 31]) are applicable only for *metallic* systems and thus not valid for our samples close to the QAH *insulating* regime (see **Fig. S9** and relevant discussion). When $(V_g - V_g^0) = -220$ V, the maximum of $\rho_{yx}^{TH}$ is ~1.65 k$\Omega$, which is much larger than the TH resistances observed in all previous studies on *metallic* systems [3, 14, 15, 16, 17, 18, 19, 20, 21].

To locate the region for the coexistence of the QAH and TH effects, we summarize the $(V_g - V_g^0)$ dependence of $\rho_{yx}(0)$, $\rho_{xx}(0)$, and $\rho_{yx}^{TH}$ at $T = 30$ mK (**Figs. 3a** and **3b**). When $-30$ V $\leq (V_g - V_g^0) \leq +40$ V, the sample exhibits a perfect QAH state, i.e. $\rho_{yx}(0)$ is fully quantized, $\rho_{xx}(0)$ and $\rho_{yx}^{TH}$ are vanishingly small. The perfect QAH state is further validated by the temperature dependence of $\rho_{xx}(0)$ and $\rho_{yx}(0)$ (**Fig. 3f**). When $(V_g - V_g^0)$ is tuned from $-30$ V to $-140$ V, the sample shows the non-perfect QAH state, where the current flows through both the dissipationless chiral edge channels and the dissipative bulk channels [32]. $\rho_{yx}(0)$ decreases from ~ $h/e^2$ to ~$0.59$ $h/e^2$, while $\rho_{xx}(0)$ increases from ~ 0 to ~ 0.87 $h/e^2$. As noted above, the existence of the QAH state for $(V_g - V_g^0) = -100$ V and $-140$ V can be further confirmed by the decrease in $\rho_{xx}(0)$ and the increase in $\rho_{yx}(0)$ with decreasing temperature (**Figs. 3d** and **3e**). The opposite temperature dependences of $\rho_{xx}(0)$ and $\rho_{yx}(0)$, together with the large $\rho_{yx}(0)/\rho_{xx}(0)$ ratio, can only be a result of the chiral edge states, thus providing conclusive evidence for the existence of the QAH state. Simultaneously, $\rho_{yx}^{TH}$ increases monotonically between $-140$ V and $-30$V for $(V_g - V_g^0)$, demonstrating the presence of the TH effect as well as the QAH effect. In other words, the chiral spin textures coexist with chiral edge states. When $(V_g - V_g^0)$ is further tuned from $-140$ V to $-220$ V, the signature of the QAH state disappears, with $\rho_{xx}(0)$ and $\rho_{yx}(0)$ showing similar temperature dependence (**Fig. 3c**), only TH effect is seen in this regime.

The QAH state persists for $(V_g - V_g^0)$ greater than $+40$ V up to $+100$ V with $\rho_{yx}(0)$ ~ $0.49$ $h/e^2$, and $\rho_{yx}(0)/\rho_{xx}(0)$ ~ 1.1. The presence of the QAH state at $(V_g - V_g^0) = +70$ V is also verified by the decrease in $\rho_{xx}(0)$ and the increase in $\rho_{yx}(0)$ with decreasing temperature (**Fig. 3g**). In this regime, a smaller $\rho_{yx}^{TH}$ as compared to the p-doped regime (i.e. $V_g < V_g^0$) is found. Therefore, for $+40$ V $< (V_g - V_g^0) \leq +100$ V, the TH effect is also concurrent with the QAH effect. When $(V_g - V_g^0)$ is further increased from $+100$ V to $+180$ V, the similar temperature dependence of $\rho_{xx}(0)$ and $\rho_{yx}(0)$ marks the disappearance of the QAH state (**Fig. 3h**), only TH effect is seen. The much smaller $\rho_{yx}^{TH}$ for $V_g > V_g^0$ indicates that the chemical potential of the system has not crossed the bulk conduction bands. As we discuss in detail below, this implies that the TH feature at $V_g > V_g^0$ is solely induced by the asymmetric potential between the top and bottom surfaces crossing the helical SSs [32, 33].

**Figure 3i** shows the $\mu_0 H$ dependence of $\rho_{yx}^{TH}$ with the maximum peak at different temperatures under $V_g = V_g^{TH, max}$. $\rho_{yx}^{TH}$ decreases gradually with temperature from 30mK and disappears at $T = 5$ K. The peak position of $\rho_{yx}^{TH}$ and the magnetic field range of the "hump" feature also monotonically decreases with increasing $T$. We summarize the peak value of $\rho_{yx}^{TH}$ (i.e. $\rho_{yx}^{TH, max}$) as a function of $(V_g - V_g^0)$ at different temperatures. The $\rho_{yx}^{TH}$ curve at each $T$ is asymmetric between $V_g < V_g^0$ and $V_g > V_g^0$. When 60 mK $\leq T \leq$ 1 K, the $\rho_{yx}^{TH}$ shows a peak when $V_g < V_g^0$ (**Fig. S7g**). This observation indicates the DM interaction strength is maximized when the chemical potential crosses the bulk valence bands.

In order to understand the experimental observations, we propose a model based on the emergence of chiral spin textures around the $\mu_0 H_c$ regime. The observed "hump" structure in $\rho_{yx}$ has been observed in a variety of noncollinear magnetic systems, particularly skyrmion systems [14, 15, 16, 17], and is regarded as the key signature for the chirality of skyrmions. However, our sample has a robust FM ground state for the occurrence of the QAH effect at low $\mu_0 H$, thus stable skyrmions are unlikely to be formed [34]. Our theoretical calculation (see **Fig. S17** and relevant discussion) confirms that the magnetic TI sample is dominated by FM states (i.e. Heisenberg magnetic coupling). Since the TH effect only occurs near the $\mu_0 H_c$ regime, this fact motivates us to consider the possible chiral spin textures during magnetization reversal. During a FM transition, magnetic domains with opposite polarization are nucleated. The presence of strong DM interaction gives rises to chiral walls at domain boundaries (**Figs. 4a** and **4b**). Net scalar chirality $Q = \sum \mathbf{S}_1 \cdot (\mathbf{S}_2 \times \mathbf{S}_3)$ is thus non-zero and leads to the TH effect. This is similar to the emergent chirality reported in Refs. [35, 36, 37], where magnetization reversal is driven by thermal fluctuations. The DM interaction is essential since chirality is caused by the canting of neighboring spins. Furthermore, since the scalar chirality $Q$ respects full rotational symmetry while the DM interaction strength $D$ breaks inversion symmetry, $Q \sim D^2$. The TH effect observed in our experiment can thus be understood qualitatively by investigating the DM interaction in magnetic TI sample.

The DM interaction is attributed to the strong spin-orbit coupling in our system. It can be computed via the spin susceptibility of a simplified model for TI films [38]. The Hamiltonian consists of two parts: the SS Hamiltonian $H_{ss}$ and the bulk state Hamiltonian $H_{QW}$ that describes the quantum well (QW) states due to the quantum confinement effect (see **Section VI** in Supplementary Information). The energy dispersions of the SS and QW bands are shown in **Fig. 4c**. For simplicity, only one set of QW bands are included, but the inclusion of more QW bands is straightforward and does not affect the qualitative physical picture. The Dirac cones of SS bands are close to the valence band top, and this is consistent with the early theoretical and experimental studies [32, 33, 39]. Electrons couple to the magnetization $\mathbf{M}$ via $H_{Zeeman} = -\mathbf{M} \cdot \mathbf{\Gamma}$, where $\mathbf{\Gamma}$ are proper $4 \times 4$ matrices for electron spin operators. The spin susceptibility $\chi_{\alpha\beta}(\alpha, \beta = x, y, z)$ is evaluated for the model Hamiltonian based on linear response theory $\chi_{\alpha\beta}(\mathbf{q}) = \frac{T}{2V} \text{Tr}[G_0(\mathbf{q} + \mathbf{k}, i\omega_m)\Gamma_\alpha G_0(\mathbf{k}, i\omega_m)\Gamma_\beta]$. The DM interaction is given by $D_\alpha(\mathbf{q}) = \varepsilon_{\alpha\beta\gamma}\chi_{\beta\gamma}(\mathbf{q})$ where $\varepsilon$ is the Levi-Civita symbol. As the system breaks mirror symmetry with respect to $xy$-plane, $\chi_{xy} = 0$, we thus focus on the off-diagonal components $\chi_{xz}$ and $\chi_{yz}$. As expected from the Moriya rule [40], $\chi_{xz}$ ($\chi_{yz}$) is linearly proportional to the momentum $q_x$ ($q_y$) (**Fig. S18**). By choosing a nonzero $q_x$ value, we calculated $\chi_{xz}$ as a function of energy for the QW states and SSs with different asymmetric potential $U$ (**Figures 4d** and **4e**) and found a nonzero $U$ is indeed required to break the inversion symmetry and induce finite $\chi_{xz}$. Note that in the real samples, in addition to the chemical potential ($\mu$), $U$ also depends on $V_g$, but as discussed in Supplementary Information, its dependence on $V_g$ is much weaker in our experiment. The SSs contribution to $\chi_{xz}$ shows a double peak structure around the charge neutral point. For the bulk QW states, $\chi_{xz}$ reveals a peak when $\mu$ lies between two spin-split (valence) bands and then drops and even changes its sign when $\mu$ crosses both spin bands. The bulk conduction band is well above the energy range of interest (-30~40 meV) and thus does not contribute. **Figure 4f** shows the total $\chi_{xz}$, which behaves similarly as $\rho_{yx}^{TH}$ in our experiment (**Fig. 3b**). The large asymmetry between the electron and hole doping sides arises because the SS is close to the valence band but well below the conduction band. At $V_g < V_g^0$, a large contribution to $\chi_{xz}$ from the bulk valence band top significantly enhances the DM interaction and $\rho_{yx}^{TH}$ in consequence. We note that including more bulk QW states in the model can further enhance the DM interaction in the hole doping regime. On the other hand, the SS contribution prevails in the $V_g > V_g^0$ regime. When $\mu$ is high above the charge neutral point, $\chi_{xz}$ vanishes, which is consistent with $\rho_{yx}^{TH}$ in the electron doping regime.

We note that $\chi_{xz}$ is non-zero near the charge neutral point, but in our experiment $\rho_{yx}^{TH}$ is not seen, particularly at low temperatures. This is because in the perfect QAH regime there are no bulk carriers to probe the chiral spin textures. This scenario is validated through numerical simulation of the TH effect for a single chiral domain wall (see **Section VII** in Supplementary Information). At higher temperatures (0.4 K ~ 3 K), bulk carriers can be thermally excited and $\rho_{yx}^{TH}$ gradually decreases (**Figs. 3i** and **S7g**). At even higher temperature (> 5 K), the vanishing of $\rho_{yx}^{TH}$ is the result of the fact that the thermal fluctuation is larger than the energy scale of DM interaction and thus destroys the chirality of magnetic domain walls.

A recent experiment has suggested an alternative interpretation of the "hump" feature of $\rho_{yx}$ observed in the SrIrO$_3$/SrRuO$_3$/SrTiO$_3$ sandwich structures [41]. Here, the SrIrO$_3$/SrRuO$_3$ and SrRuO$_3$/SrTiO$_3$ interfaces have opposite signs of the AH contribution and different $\mu_0H_c$. In their sandwich sample, the antiferromagnetic alignment configuration between the two magnetic interfaces will induce a larger $\rho_{yx}$ than the FM alignment. This can lead to the TH effect-like "hump" structure. However, this scenario is unlikely to occur in our samples for several reasons: First, $\rho_{yx}$ with a negative sign together with a significantly enhanced $\mu_0H_c$ has *never* been observed in Cr-doped (Bi,Sb)$_2$Te$_3$ samples. Second, as a control experiment, we grew 5QL V-doped TI (Bi,Sb)$_2$Te$_3$ on top of the 3-5-3 sandwich sample and used the exchange coupling to increase the $\mu_0H_c$ of the top Cr-doped TI layer. This structure configuration favors the formation of the antiferromagnetic alignment between top and bottom Cr-doped TI layers. The "hump" feature in this control sample, however, disappears rather than being enhanced (**Fig. S16**). Finally, our magnetization measurements show no additional step near $\mu_0H_c$ in our samples (**Fig. S3**).

To summarize, we fabricated magnetic TI sandwich heterostructures and observed the concurrence of the QAH and TH effects by applying an electrostatic gating voltage. This concurrence indicates an interplay between the chiral edge states and the chiral spin textures in magnetic TI heterostructures that deserves further experimental and theoretical studies. The chiral spin texture associated with the TH effect can be utilized to record the spin information [34], and this spin information can be transferred through the chiral edge channels of the QAH effect. The marriage of chiral edge states and chiral spin textures potentially opens the door for further explorations of proof-of-concept magnetic TI-based spintronic and electronic devices.

## Methods

**MBE growth of TI sandwich heterostructure.**

The magnetic TI sandwich heterostructure growth was carried out using a Veeco/Applied EPI MBE system with a vacuum ~ $2\times10^{-10}$ mbar. The heat-treated insulating $SrTiO_3$ (111) substrates were outgassed at ~530 °C for 1 hour before the growth of the TI sandwich heterostructures. High-purity Bi (5N), Sb (6N), Cr (5N) and Te (6N) were evaporated from Knudsen effusion cells. During growth of the TI, the substrate was maintained at ~240 °C. The flux ratio of Te per (Bi + Sb) was set to be >10 to prevent Te deficiency in the samples. The magnetic or nonmagnetic TI growth rate was at ~0.25 QL/min. Each layer of the sandwich heterostructure was grown with the different Bi/Sb ratio by adjusting their K-cell temperatures to tune the chemical potential close to its charge neutral point. Finally, to avoid possible contamination, a 10 nm thick Te layer is deposited at room temperature on top of the sandwich heterostructures prior to their removal from the MBE chamber for *ex-situ* transport and other characterization measurements.

**Hall-bar device fabrications.**

The magnetic TI sandwich heterostructures grown on 2 mm × 10 mm heated-treated insulating $SrTiO_3$ (111) were scratched into a Hall bar geometry using a computer-controlled probe station. The effective area of the Hall bar device is ~ 1 mm × 0.5 mm. The electrical ohmic-contacts for transport measurements were made by pressing indium spheres on the Hall bar. The bottom gate electrode was prepared through an indium foil on the back side of the $SrTiO_3$ substrate.

**Transport measurements.**

Transport measurements were conducted using both a Quantum Design Physical Property Measurement System (PPMS; 2 K, 9 T) and a Leiden Cryogenics dilution refrigerator (10 mK, 9 T) with the magnetic field applied perpendicular to the film plane. The bottom gate voltage was applied using a Keithley 6430. The excitation currents in the DC PPMS measurements ($\geq 2$ K) is 1 µA. We used a PicoWatt AVS-47 AC resistance bridge to conduct the dilution refrigerator measurements ($< 2$ K) with a low excitation current (1 nA) to suppress heating effect. The results reported here have been reproduced on 2 samples measured in the dilution refrigerator and more than 10 samples measured in PPMS. All the transport results shown here were anti-symmetrized as a function of the magnetic field. More transport results are found in the Supplementary Information.

**Theoretical calculations.**

The QW Hamiltonian is $H_{QW} = \varepsilon_0(\mathbf{k}) + N(\mathbf{k})\tau_z + A(k_y\sigma_x - k_x\sigma_y)\tau_x + U\tau_x$, where Pauli matrices $\sigma$'s stand for spins and $\tau$'s stand for two orbitals. $\varepsilon = C_0 + C_1 k^2$ and $N = N_0 + N_2 k^2$. Different sets of QW states differ by different $C_0$ and $N_0$ values. Dispersion in **Fig. 4c** takes values $C_0 = 0.145$ eV, $C_2 = 10.0$ eV $\cdot$ Å$^2$, $N_0 = -0.18$ eV, $N_2 = 15.0$ eV $\cdot$ Å$^2$, $A = 3.0$ eV $\cdot$ Å, and $U = 0.02$ eV. Coupling of the QW electrons to magnetization $\mathbf{M}$ is simply $H_{Zeeman}^{QW} = -\mathbf{M} \cdot \boldsymbol{\sigma}$. On the other hand, the SSs have the Hamiltonian $H_{SS} = v_F(k_y\sigma_x - k_x\sigma_y)\tau_z + U\tau_z + m_0\sigma_x$, where $\sigma$'s again stand for spins but $\tau$'s stand for two surfaces instead. $U$ is the asymmetric potential applied to two surfaces. The coupling to magnetizations $\mathbf{M}^t$ and $\mathbf{M}^b$ on top and bottom surfaces, respectively, is $H_{Zeeman}^{SS} = \mathbf{M}^t \cdot \boldsymbol{\sigma}(1 + \tau_z)/2 + \mathbf{M}^b \cdot \boldsymbol{\sigma}(1 - \tau_z)/2$. In **Fig. 4c**, we use $m_0 = 0.005$ eV, $v_F = 3.0$ eV $\cdot$ Å, and $U = 0.02$ eV.

**Acknowledgments**

The authors would like to thank X. D. Xu, B. H. Yan, H. Z. Lu and W. D. Wu for helpful discussions. D. X. and N. S. acknowledge support from ONR grant (N-000141512370) and Penn State 2DCC-MIP under NSF grant DMR-1539916. J. H. S. and M. H. W. C. acknowledge the support from NSF grant DMR-1707340. C. X. L. acknowledges the support from ONR grant (N00014-15-1-2675 and renewal No. N00014-18-1-2793). D. A. and J. D. Z. acknowledge the support from DOE grant (DE-SC0016424). C. Z. C. acknowledges the support from Alfred P. Sloan Research Fellowship and ARO Young Investigator Program Award (W911NF1810198). Support for transport measurements and data analysis is provided by DOE grant (DE-SC0019064).

**Additional information**

Supplementary information is available in the online version of the paper. Reprints and permissions information is available online at www.nature.com/reprints.

Correspondence and requests for materials should be addressed to N. S., M. H. W. C. or C. Z. C.

**Competing financial interests**

The authors declare no competing financial interests.

**Figures and figure captions**

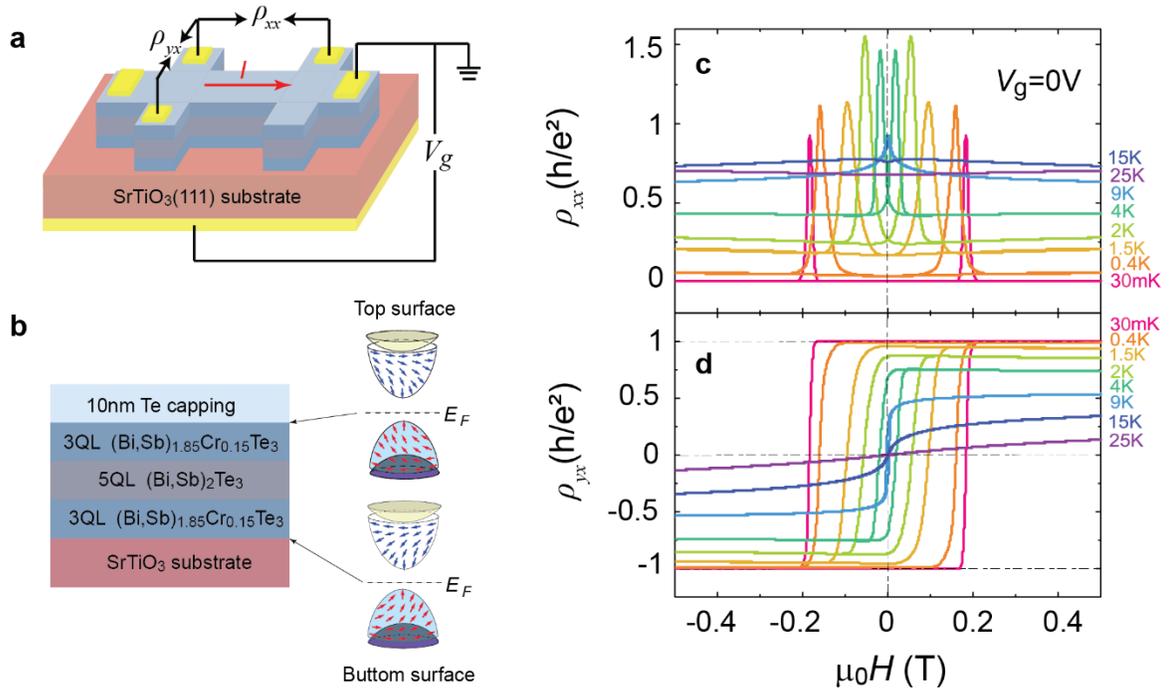

**Figure 1 | QAH effect in TI sandwich heterostructures.** (a) Schematic of the field-effect transistor device with a bottom gate ($V_g$) used in transport measurements. The electrical contacts on the Hall bar and the back-gate contact are made by pressed indium dots. The 0.5 mm SrTiO$_3$(111) substrate is used as the dielectric layer for the bottom gate. (b) Schematic of the magnetic TI sandwich heterostructure. The total thickness of the sample is 11QL. When $T < T_C$, an exchange gap opens at the Dirac points of the top and bottom surfaces. Blue (red) arrows in the gapped Dirac surface states represents the spin orientations. (c, d) Magnetic field $\mu_0 H$ dependence of the longitudinal resistance $\rho_{xx}$ (c) and the Hall resistance $\rho_{yx}$ (d) at $V_g = 0$ V. At $T = 30$ mK and $V_g = 0$ V, the quantized $\rho_{yx}$ and the vanishing $\rho_{xx}$ suggest this sandwich sample is in the QAH state.

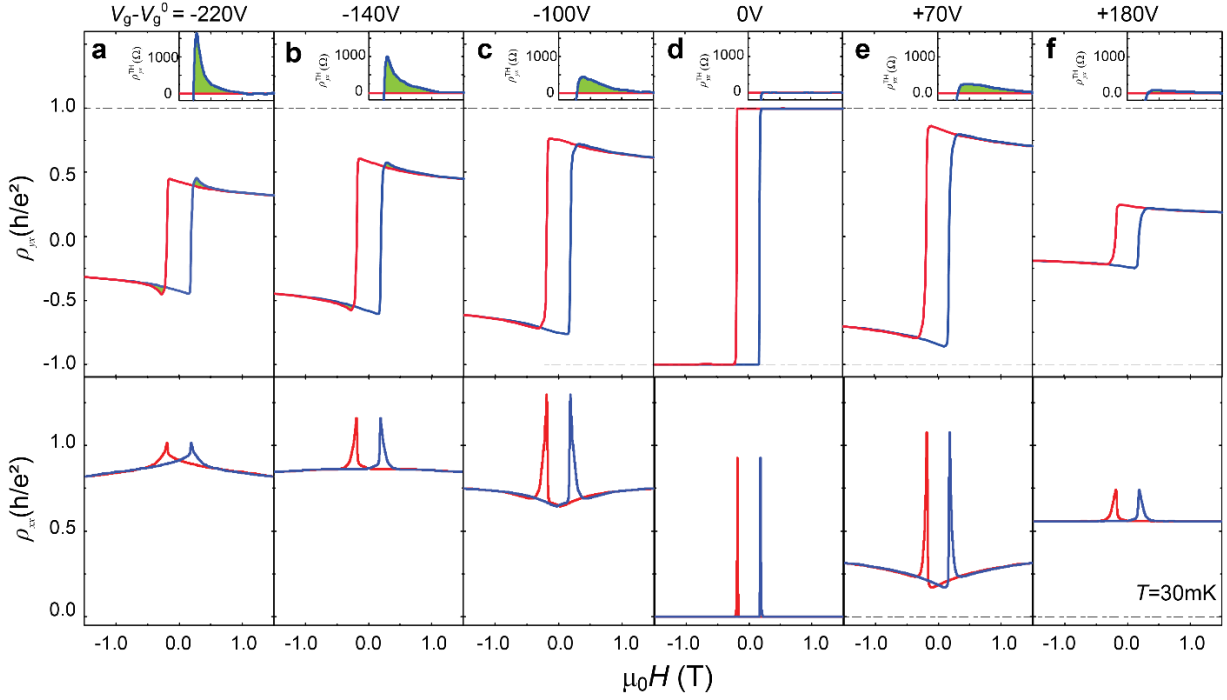

**Figure 2 | Gate-induced TH effect in TI sandwich heterostructures.** (a-f) Magnetic field $\mu_0 H$ dependence of the Hall resistance $\rho_{yx}$ (top) and the longitudinal resistance $\rho_{xx}$ (bottom) at different gates ($V_g - V_g^0$). The sample shows a perfect QAH state when $V_g = V_g^0 = +20$ V. When $V_g$ is tuned away from $V_g^0$, $\rho_{yx}$ deviates from the quantized value (i.e., $h/e^2$) and shows a "hump" feature shaded in green which is known as the TH effect. Insets of (a-f) show the TH resistance $\rho_{yx}^{TH}$, which is subtracted using the following method: the offset resistance of $\rho_{yx}$ when the external $\mu_0 H$ is swept upward and downward. Blue (red) curve represents the process for increasing (decreasing) $\mu_0 H$.

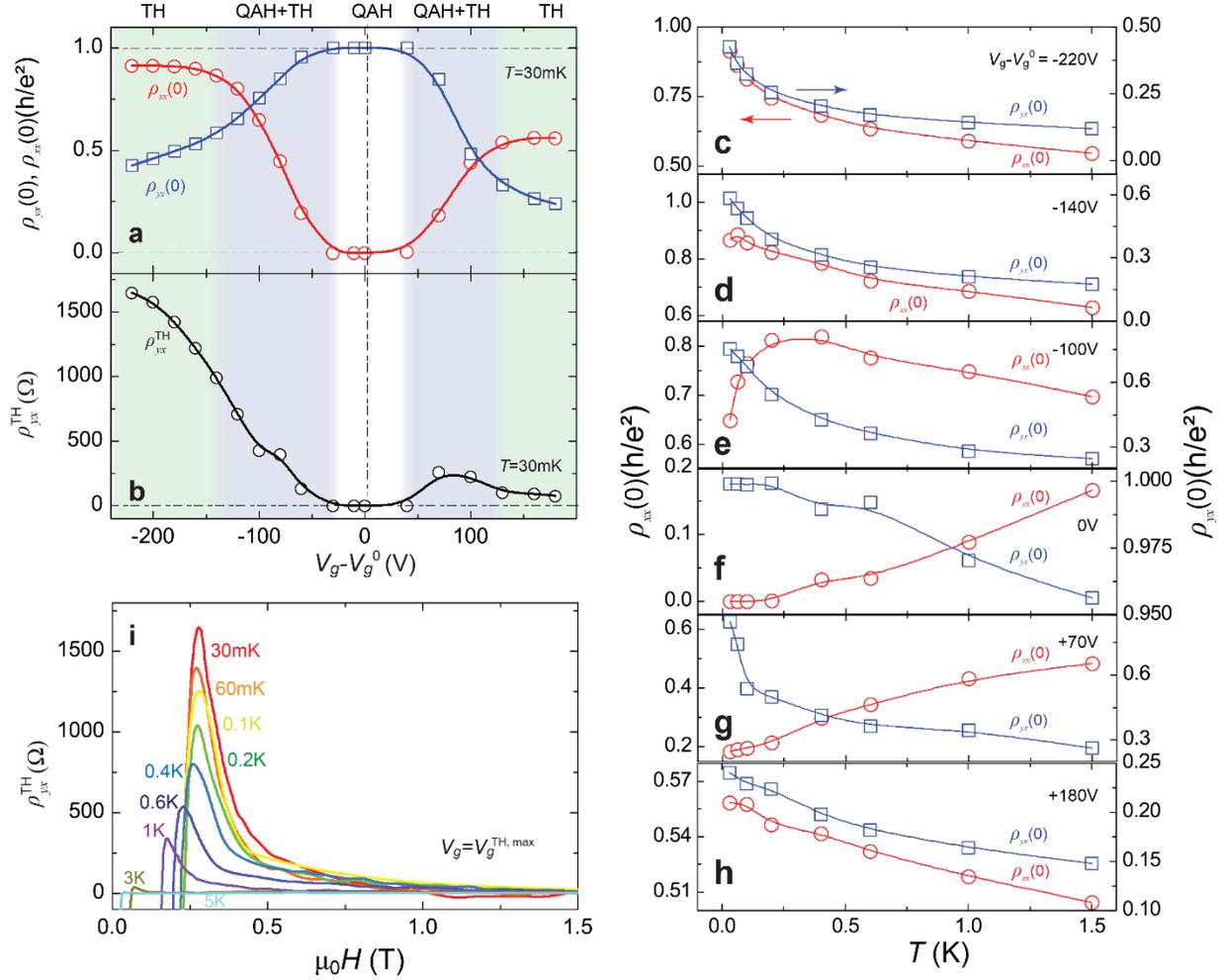

**Figure 3 | Concurrence of the QAH and TH effect in TI sandwich heterostructures.** (a) Gate dependence of the Hall resistance $\rho_{yx}(0)$ (empty blue squares) and the longitudinal resistance $\rho_{xx}(0)$ (empty red circles) at zero magnetic field. (b) Gate dependence of the TH resistance $\rho_{yx}^{TH}$. The regions of concurrence of the QAH and TH effects are shaded in light blue. (c-h) Temperature dependence of the Hall resistance $\rho_{yx}(0)$ (empty blue squares) and the longitudinal resistance $\rho_{xx}(0)$ (empty red circles) for different gates ($V_g - V_g^0$). (i) $\mu_0H$ dependence of $\rho_{yx}^{TH}$ for different $T$ at $V_g=V_g^{TH,\ max}$. $\rho_{yx}^{TH}$ decreases with increasing temperature. $\rho_{yx}^{TH}$ is 1.65 KΩ at $T$ = 30 mK and disappears at $T$ = 5K.

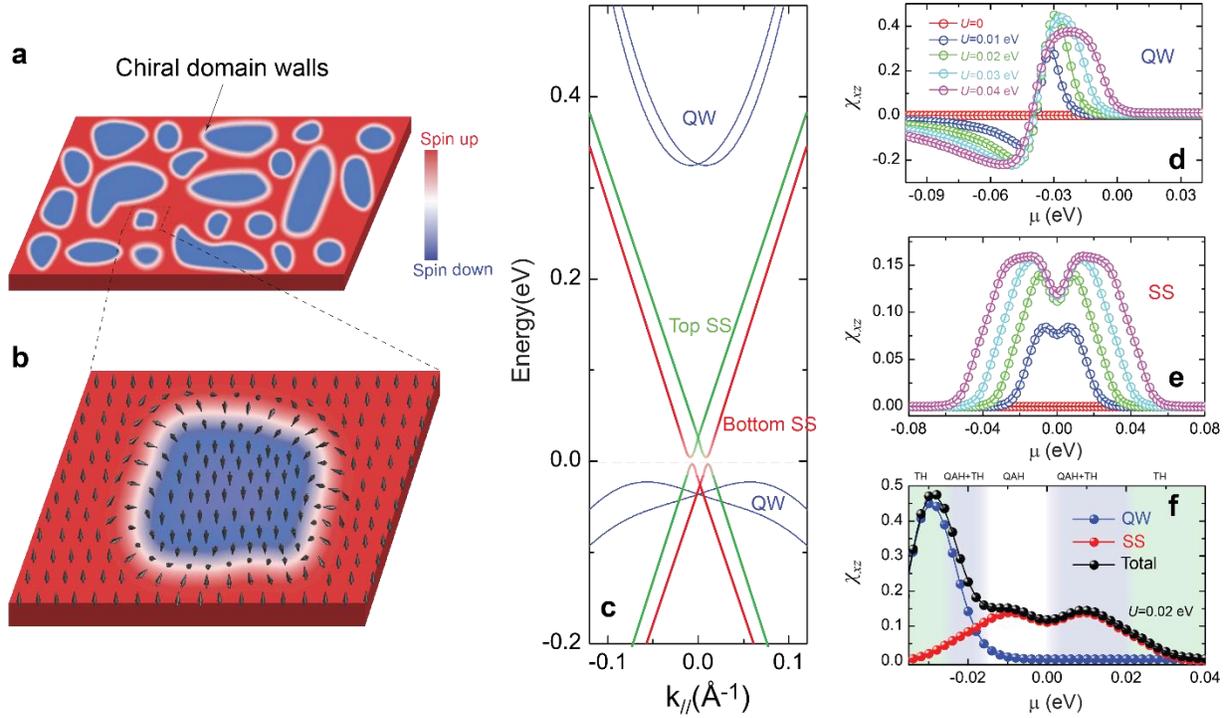

**Figure 4 | Chiral domain walls and theoretical interpretations of the appearance of the TH effect.** (a) The formation of the chiral domain walls during magnetization reversals. (b) A magnified view depicting the spin distribution of the chiral domain wall. (c) The energy dispersions of the surface states (SS) and bulk quantum well (QW) bands in magnetic TI. The top SS, the bottom SS, and the bulk QW are shown in green, red, and blue, respectively. (d-e) $\chi_{xz}$ as a function of chemical potential $\mu$ for the QW and SS states, respectively, under different asymmetric potentials $U$. (f) The QW contribution to $\chi_{xz}$, SS contribution to $\chi_{xz}$, and the total $\chi_{xz}$ in the magnetic TI sandwich heterostructures when $U$=0.02 eV. $q_x = 0.005$ Å$^{-1}$ and $q_y = 0$ in (d-f).